\documentclass{aastex}
\usepackage{emulateapj5,psfig}
\usepackage{graphicx}
\usepackage{times}
\def\ltsima{$\; \buildrel < \over \sim \;$}
\def\lsim{\lower.5ex\hbox{\ltsima}}
\def\gtsima{$\; \buildrel > \over \sim \;$}
\def\gsim{\lower.5ex\hbox{\gtsima}}

\newcommand{\be}{\begin{equation}}
\newcommand{\en}{\end{equation}}
\newcommand{\ergs}{\rm \ erg \; s^{-1}}

\def\cmdue {\rm \ cm^{-2}}

\def\msole {~M_{\odot}}

\def\deg {^\circ}

\begin{document}

\title{An XMM-Newton study of the 401 Hz accreting pulsar SAX J1808.4--3658 in
quiescence} 
\shortauthors{Campana et al.}

\author{S.~Campana\altaffilmark{1}, L. Stella\altaffilmark{2}, 
F. Gastaldello\altaffilmark{3}, S. Mereghetti\altaffilmark{3},
M. Colpi\altaffilmark{4}, G.L. Israel\altaffilmark{2},
L. Burderi\altaffilmark{2}, T. Di Salvo\altaffilmark{5},
R.N. Robba\altaffilmark{6}}

\altaffiltext{1}{INAF-Osservatorio Astronomico di Brera, Via Bianchi 46, I--23807
Merate (Lc), Italy}

\altaffiltext{2}{INAF-Osservatorio Astronomico di Roma,
Via Frascati 33, I--00040 Monteporzio Catone (Roma), Italy}

\altaffiltext{3}{CNR-IASF, Istituto di Astrofisica Spaziale e Fisica Cosmica
Sezione di Milano ``G.Occhialini'', Via Bassini 15, I--20133 Milano, Italy}

\altaffiltext{4}{Dipartimento di Fisica G. Occhialini, Universit\`a di
Milano Bicocca, Piazza della Scienza 3, I--20126 Milano, Italy}

\altaffiltext{5}{Astronomical Institute A. Pannekoek, University of Amsterdam
and Center for High-Energy Astrophysics, Kruislaan 403, NL 1098 SJ Amsterdam,
The Netherlands}  

\altaffiltext{6}{Dipartimento di Scienze Fisiche ed Astronomiche,
Universit\`a di Palermo, Via Archirafi 36, I--90123 Palermo, Italy}

\email{campana@merate.mi.astro.it}

\begin{abstract}
SAX J1808.4--3658 is a unique source being the first Low Mass X--ray Binary
showing coherent pulsations at a spin period comparable to that of millisecond 
radio pulsars. 
Here we present an XMM-Newton observation of SAX J1808.4--3658 in quiescence,
the first which assessed its quiescent luminosity and spectrum
with good signal to noise. 
XMM-Newton did not reveal other sources in the vicinity of SAX 
J1808.4--3658 likely indicating that the source was also detected by previous
BeppoSAX and ASCA observations, even if with large positional and flux
uncertainties. We derive a 0.5--10 keV unabsorbed luminosity of
$L_X=5\times10^{31}\ergs$, a relatively low value compared with other neutron
star soft X--ray transient sources. At variance with other soft X--ray
transients, the quiescent spectrum of SAX J1808.4--3658 was dominated by a hard
($\Gamma\sim 1.5$) power law with only a minor contribution ($\lsim 10\%$)
from a soft black body component. If the power law originates in 
the shock between the wind of a turned-on radio pulsar and matter outflowing
from the companion, then a spin-down to X--ray luminosity 
conversion efficiency of $\eta\sim 10^{-3}$ is derived; this is in line 
with the value estimated from the eclipsing radio pulsar PSR J1740--5340. 
Within the deep crustal heating model, the faintness of the blackbody-like 
component indicates that SAX J1808.4--3658 likely hosts a massive neutron
star ($M\gsim1.7\msole$).
\end{abstract}

\keywords{accretion, accretion disks --- binaries: close --- star: individual
(SAX J1808--3658) --- stars: neutron}

\section{Introduction}

Neutron star Soft X--ray Transients (SXRTs) when in outburst closely resemble
persistent Low Mass X--ray Binaries (LMXRBs). In the last few years it has
become clear that SXRT sources form a rather inhomogeneous class (see Campana et
al. 1998a for a review) comprising sources with well defined outbursts
as well as sources with long on/off activity periods.
 Moreover, sources displaying bright outbursts with peak X--ray
luminosities $L_X\sim 10^{37}-10^{38}$~erg~s$^{-1}$ appear to be different
from sources showing only faint outbursts reaching
$L_X\sim 10^{36}-10^{37}$~erg~s$^{-1}$, especially in the Galactic center
region (Heise et al. 1998; King 2000; in't Zand 2001a).

A major leap forward came with the discovery of SAX J1808.4--3658,
a bursting SXRT reaching a maximum luminosity of $\sim 2\times 10^{36}\ergs$
(for a distance of 2.5 kpc; in't Zand al. 2001b). In April 1998 the source
resumed activity and RXTE observations revealed coherent $\sim 401$~Hz
pulsations, the first detected in the persistent emission of a
neutron star LMXRB. These testify to the presence of magnetic
polar cap accretion onto a fast rotating magnetic neutron star (Wijnands \&
van der Klis 1998; Chakrabarty \& Morgan 1998).
The inferred magnetic field strength of SAX~J1808.4--3658 is in the
$10^8-10^9$~G range (see Psaltis \& Chakrabarty 1999), providing convincing
evidence for the long suspected LMXRB-millisecond pulsar connection.

SXRTs spend much of their time in quiescence.
The origin of the quiescent X--ray emission is still uncertain. In the last
few years several sources have been studied in detail; the picture
that emerged is that the quiescent spectrum consists of a soft component
plus a high energy excess.
The former component is often fit with a blackbody spectrum with an equivalent
radius of 1--2~km and temperatures in the 0.1--0.3~keV range or with a neutron
star atmosphere model with an equivalent radius consistent with the entire
neutron star surface and slightly smaller temperatures (Brown et al. 1998;
Rutledge et al. 2000). The high energy component is well represented by a
power law with photon index $\Gamma\sim 1-2$ (Asai et al. 1996, 1998; Campana et
al. 1998b, 2000).  
In all sources observed so far the quiescent luminosity ranges between
$10^{32}-10^{33}\ergs$, indicating a clear difference with black hole
transients in quiescence that have a lower X--ray luminosity
(Garcia et al. 2001; Campana \& Stella 2000).
Flux and spectral variability have been reported in Aql X-1 and KS
1731--260 during quiescence (Campana et al. 1997; Rutledge et al. 2002;
Wijnands et al. 2002a). 
This poses severe limitations on the emission mechanisms responsible for the
quiescent luminosity.

Among neutron star SXRTs, SAX J1808.4--3658 stands out for having, while in
outburst, a magnetosphere and, of course, an accurately measured spin
period. These characteristics make the source very well suited for testing the 
predictions of models for the quiescent emission in which the presence of a
sizeable magnetic field plays a crucial role (Stella et al. 1994).
SAX J1808.4--3658 was detected in quiescence (Stella et al. 2000;
Dotani et al. 2001; Wijnands et al. 2001) though with large uncertainties.
Here we report on an XMM-Newton observation of SAX~J1808.4--3658 in
quiescence, the first to detect the source with a good signal to noise
ratio.

\section{Data analysis}

SAX J1808.4--3658 was observed on March 24 2001 with XMM-Newton EPIC,
consisting of two metal oxide semiconductor (MOS) cameras (Watson et al. 2000)
and one pn camera (Str\"uder et al. 2000).
Medium filters were used for the MOS cameras and the thin filter for
the pn camera. The pn camera was operated in timing mode in order to search  
for pulsations in case the source were sufficiently bright. SAX J1808.4--3658
turned out to be faint (see below), preventing meaningful searches for pulsed
emission. We do not discuss the pn camera data in the following and
concentrate on data from MOS cameras that were operated in full frame mode
with a read-out time of 2.6 s. 

We extracted the event file starting from the raw data using the
Standard Analysis System (SAS) version 5.3.0. Data were manually
screened to remove any remaining bright pixels or hot columns. Periods in
which the background was high because of soft proton flares were excluded
using an intensity filter: we rejected all events accumulated in the external
CCDs (2--7) when the total count rate exceeded 15 counts in 100 s in the
10--12.4 keV band for the MOS1 and MOS2, independently. 
We obtained a net exposure time of 33.2 ks and 34.4 ks for MOS1
and MOS2, respectively (this is slightly different from what reported in
Campana et al. 2002, due to the use of standard processed files with SAS
5.0.3). Spectra were accumulated separately for MOS1 and MOS2. Event grades
higher than 12 were filtered out. Photons were  
grouped within {\tt XMMSELECT} to the nominal MOS resolution of 15 eV.

\subsection{Source position}

Previous observations aiming at determining the position of SAX J1808.4--3658
were performed with BeppoSAX (Stella et al. 2000; Wijnands et al. 2002b) and
ASCA (Dotani, Asai \& Wijnands 2000; see Table 1), but due to the limited
angular resolution of these satellites the resulting picture was quite
confused. In particular it was unclear whether the source revealed at a low
flux of $\sim 10^{-13}\ergs\cmdue$ and with large positional uncertainties was
indeed SAX J1808.4--3658 (see Fig. 1 and Tab. 1). In fact, Wijnands et al.
(2002b) claimed that SAX J1808.4--3658 was only detected in the ASCA data of
September 1999, while neither BeppoSAX observations detected the millisecond
pulsar but rather revealed a nearby source, SAX J1808.6--3658. 

The deep image we obtained with EPIC allowed us to clarify this
confused picture by revealing several faint sources in the region.
Source detection was obtained following the prescriptions of Baldi et al.
(2002). This revealed 20 sources in the central part ($12'\times9'$) of the
field of view. The brightest one was clearly coincident with
the radio position of SAX J1808.6--3658. A few other sources were contained in
the error regions of sources previously seen with ASCA and in the longer
BeppoSAX observation (see Fig. 1).

A faint source is visible about $2'$ East of SAX J1808.4--3658 (RA(J2000):
18$^{\rm h}$ 08$^{\rm m}$ 36$^{\rm s}$.6; DEC(J2000):--36$\deg$ $58'$
$01''$). This is within the error region of the source detected by Wijnands
et al. (2002b) in the March 2000 BeppoSAX observation. If it had the same
luminosity that we measure with EPIC (a count rate a factor $\sim 5$ lower
than that of SAX J1808.4--3658), it was probably too faint to be detected by
BeppoSAX, but a small contamination from this source could perhaps explain why
all the error regions for SAX J1808.4--3658 derived with low angular
resolution instruments are systematically centered to the East of the radio
position. A further bias is provided by a $2.9'$ to the brighter source
East (RA(J2000): 18$^{\rm h}$ 08$^{\rm m}$ 42$^{\rm s}$.2;
DEC(J2000):--36$\deg$ $58'$ $44''$) which has a comparable flux to SAX
J1808.4--3658 and went undetected in previous observations (Wijnands et
al. 2002b). Furthermore, given the source density revealed in this field 
by XMM-Newton, the probability of finding by chance at least one source within
the BeppoSAX $90\%$ error circle of SAX J1808.6--3658 is $\sim 70\%$. It
appears to be unlikely that this faint source was much brighter and at the
same time SAX J1808.4--3658 much fainter just during the BeppoSAX observations
and conclude that SAX J1808.4--3658 was the source detected both in March 1999
and March 2000. 

\subsection{X--ray spectrum and luminosity}

In total we collected 91 and 77 counts with the MOS1 and MOS2 detectors,
respectively, within a $10''$ radius circle centered on the position of SAX
J1808.4--3658. This corresponds to about $55-60\%$ of the source flux for 
energies between 1--5 keV (extending the extraction region would just decrease 
the signal to noise ratio, due to the background and source
faintness). The background was extracted from an annular region with inner and
outer radii of $0.5'$ and $2.5'$ (excluding contaminating sources),
respectively. Within the SAX J1808.4--3658 extraction region, about $15-20\%$
of the counts are to be attributed to the local background. 

The MOS1 and MOS2 spectra were rebinned to contain at least 15 photons per channel.
The spectral analysis was carried out with XSPEC v.11.1.0. We adopted the latest
on-axis standard response matrices ({\tt m1\_medv9q20t5r6\_all\_15.rsp} and
{\tt m2\_medv9q20t5r6\_all\_15.rsp})
and correct the fluxes for the fraction of photons collected within the
extraction radius. The spectral analysis was carried out in the 0.3--7
keV energy range (Kirsch private communication and 2002). Given that the
cross calibration between MOS1 and MOS2 agrees within $5\%$ (Kirsch 2002) we
did not include a constant for the two MOS instruments.
The statistics was relatively poor and the spectrum could be
well fit by an absorbed power law (or a bremsstrahlung) model (see Table 2
and Fig. 2). 
A power law model gave a better fit ($\chi^2_{\rm red}=1.1$) with
$\Gamma=1.5^{+0.2}_{-0.3}$ ($90\%$ confidence level) and a column density of
$N_H=0.3\times10^{21}\cmdue$ ($<1.1\times10^{21}\cmdue$ $90\%$ c.l., we used
here the Wisconsin cross sections in the {\tt wabs} model).
We note that a model with the column density fixed at the value derived from
the source data in outburst (e.g. Gilfanov et al. 1998) and consistent with
the galactic value of $1.3\times 10^{21}\cmdue$, provides also a good fit with
$\Gamma=1.8\pm0.3$ and $\chi^2_{\rm red}=1.2$ (with a null hypothesis
probability of $20\%$). 
The unabsorbed 0.5--10 keV luminosity of SAX J1808.4--3658 was
$5\times10^{31}\ergs$ for the power law model and a (revised) distance of 2.5
kpc (in't Zand et al. 2001b).
Single component black body or neutron star atmosphere models (Gaensicke,
Braje \& Romani 2002) did not provide an adequate fit to the data with
$\chi^2_{\rm red}>3$.

Given that the best studied quiescent SXRTs display a
soft black body component plus a hard power law tail we consider also this
model (even though strictly not required by the data). 
In this case we fixed again the column density to $1.3\times
10^{21}\cmdue$. The power law fit alone could adequately describe 
the data so that we first fixed the best fit power law and derived constraints
on the black body component. SXRTs in quiescence display black body
temperatures ranging between 0.1--0.3 keV (e.g. Rutledge et al. 2000). By
using this range we derived a $90\%$ upper limit to the normalization of the
black body component and in turn on the (unabsorbed, bolometric) soft black
body flux of $\sim 2\times 10^{-15}\ergs\cmdue$. This corresponds to a factor
of 15 less than the power law luminosity in the 0.5--10 keV band. If we repeat
the same exercise with a neutron star atmosphere model (Gaensicke, Braje \&
Romani 2002) we derive a factor of 30 less luminosity in the soft component.
Alternatively, by fixing the normalisation ratio of the black body and power
law components as observed in the best studied quiescent SXRTs (e.g. Campana
et al. 1998b, even if recent observations on Aql X-1 challange this
picture, Rutledge et al. 2002) we obtained a very low black body temperature.
By fixing also the column density to $1.3\times 10^{21}\cmdue$ we succeeded in
obtaining a good fit $\chi^2_{\rm red}=1.2$ with a black body temperature of
$k\,T=0.16^{+0.04}_{-0.05}$ keV and a power law $\Gamma=1.5\pm0.5$. The source 
flux is in any case low $2.5\times 10^{31}\ergs$. With a neutron star
atosphere model we obtain an equally good fit ($\chi^2_{\rm red}=1.2$) with a
temperature $k\,T=0.09^{+0.03}_{-0.03}$ keV and an emitting radius
$R<1.1$ km.

\section{Discussion}

The high throughput and good angular resolution of XMM-Newton provided the first
firm determination of the quiescent luminosity of SAX J1808.4--3658. This is
lower by a factor of $\sim 2$ than previous best fit estimates. 
However, once the large uncertainties in the spectral parameters are taken
into account, a fairly constant luminosity is inferred since March 1999 at a
level of $5\times 10^{31}\ergs$ (at a distance of 2.5 kpc; note that the
reduced $\chi^2$ of a fit with a constant is 1.8, $15\%$ null hypothesis
variability, see Fig. 3). We conclude that the source variability issue
raised by Dotani et al. (2000) is questionable.
The source luminosity is a factor of two lower than that usually
measured in SXRTs in quiescence; this ranges between $10^{32}-10^{33}\ergs$
(e.g. Campana et al. 1998a; Campana \& Stella 2000; Garcia et al. 2001). 
For a distance of 4 kpc, however, it would be fully consistent with what
usually observed in other SXRTs. 

More interestingly, if we consider the spectral fit with a power law component
plus a black body, the soft X--ray component comprises only a small part 
($\lsim 10\%$) of the total luminosity, whereas in the great majority of SXRTs
it accounts for about half of the luminosity in the 0.5--10 keV energy band
(e.g. Campana et al. 1998b, 2000; see however Rutledge et al. 2002). 
The soft component is usually ascribed to the cooling of the neutron star surface
powered by the deep nuclear heating that the neutron star receives during each
outburst (Brown et al. 1998; Rutledge et al. 2000; Colpi et al. 2001; see also
Campana et al. 1998a). RXTE All Sky Monitor (ASM) provides us with a
continuous monitoring of the high level activity of SAX J1808.4--3658.
 From these data we can extrapolate a mean mass transfer rate of $\sim 5\times
10^{-12}\msole$ yr$^{-1}$, in line with previous estimates (Bildsten \&
Chakrabarty 2001). This mean mass inflow rate translates into a soft
quiescent luminosity of $10^{32}\ergs$ within the 0.5--10 keV energy band
from deep crustal heating (Brown et al. 1998; Colpi et al. 2001).
This is a factor of $\sim 10$ higher than observed. 
Given the fact that predictions from deep nuclear heating are quite robust,
one is led to conclude that an additional source of cooling is present.
A simple and well known solution is when the direct Urca process is allowed
in the neutron star core and, in turn, neutrino cooling does affect the
neutron star thermal evolution (Colpi et al. 2001). This can occur only for
massive neutron stars with masses higher than $1.7-1.8\msole$.
If this interpretation were correct the neutron star of SAX J1808.4--3658 
have to be fairly massive, in agreement with accretion spin-up scenarios. 

The main contribution to the quiescent luminosity of SAX J1808.4--3658 appears
to derive from the power law component. A pure propeller contribution is
ruled out since this mechanism should stop operating at a luminosity of about 
$10^{33}\ergs$, with the turing on of a radio pulsar (e.g. Campana et
al. 1998a, 1998b).
One interpretation for this power law relies on the emission at the
shock front between the relativistic wind of a radio pulsar and matter
outflowing from the companion star (see the discussion in Stella et al. 2000).
Being the first SXRT for which the presence of 
a sizable magnetic field is unambiguously established, 
SAX J1808.4--3658 can be used to infer the efficiency $\eta$ with which 
spin-down power is converted into 0.5--10 keV luminosity. 
We derive $\eta\sim 5\times 10^{-3}\,B_8^2$, where $B_8=B/10^8\, {\rm G}$ is 
neutron star magnetic field (as derived from the accretion luminosity at the 
propeller onset, Gilfanov et al. 1998; see also Psaltis \& Chakrabarty 1999).  

Recently, an ideal laboratory for studying the shock emission has been
discovered: this is PSR J1740--5340, a 3.7 ms radio pulsar in a 
32.5 hr orbit around a Roche lobe filling main sequence companion, 
which display partial and total eclipsed over a wide range of orbital phases 
(D'Amico et al. 2001; Ferraro et al. 2001; Burderi, D'Antona \& Burgay 2002). 
The X--ray luminosity of  PSR J1740--5340 measured by Chandra 
($8\times 10^{30}\ergs$, 0.5--2.5 keV range, unabsorbed; Grindlay et al. 2001)
implies $\eta\sim 10^{-4}$. 
Taking this system as a prototype for modelling the shock emission power
(which is likely less efficient since the radio pulsar is not  completely
engulfed) and scaling the efficiency as the square of the orbital separation,
one infers $\eta\sim 3\times 10^{-3}$, well consistent with the value above,
given the uncertainties involved.

\begin{acknowledgements}
We thank A. Baldi for the source detection algorithm and A. De Luca for 
the image hot pixels cleaning procedure. 
We acknowledge useful comments by an anonymous referee which helped improving
the paper.
\end{acknowledgements}

\vskip 3truecm

\begin{table}[!h]
\caption{SAX J1808.4--3658 positions.}
\begin{tabular}{cccccc}
\hline
Obs. date& Satell.&Exp. time& Error radius        & Distance$^*$ & Refs.$^{\dag}$ \\
\hline
Mar. 1999$^+$&BSAX& 20 ks   &$1.50'$ ($90\%$)&$1.43'$       & 1 \\
Sep. 1999& ASCA   & 61 ks   &$0.24'$ ($68\%$)&$0.19'$       & 2 \\
Mar. 2000& BSAX   & 89 ks   &$0.55'$ ($68\%$)&$1.63'$       & 3 \\
Mar. 2001& XMM    & 33 ks   &$0.10'$ ($90\%$)&$0.04'$       & 4 \\
\hline
\end{tabular}

\noindent $^*$ Distance of the detected source from the radio position of SAX
J1808.4--3658.

\noindent $^\dag$ 1: Stella et al. (2000); 2: Dotani et al. (2000); 3:
Wijnands et al. (2002); 4: this paper.

\noindent $^+$ Source position derived is affected by a large uncertainty due
to source faintness and unfavorable roll angle resulting in a systematic
uncertainty of about $35''$ on top of the intrinsic positional error (Perri \&
Capalbi 2002). 

\end{table}

\begin{table*}[!htb]
\caption{Spectral models and related luminosities.}
\begin{tabular}{cccccc}
\hline
Model          & $N_H$           &$\Gamma$/$k\,T$    &$\chi^2_{\rm red}$&Flux& $L_X$ \\
               &($10^{21}\cmdue$)&                   &                  &($10^{-14}$ cgs) & ($10^{31}\ergs$)\\
\hline
Black body$^+$ &$0.0^{+0.3}$     &$0.52_{-0.08}^{+0.11}$ keV& 3.2       & 1.5& $2.2$ \\
Power law      &$0.3^{+0.7}$     &$1.5_{-0.3}^{+0.2}$&   1.1            & 2.9& $4.6$ \\
Bremsstrahlung &$0.0^{+0.9}$     &$15.5_{-10.4}$ keV &   1.1            & 2.8& $4.4$ \\
NS atmosphere$^*$&$0.4^{+1.1}$   &$0.27_{-0.02}^{+0.0}$ keV& 4.4        & 1.1& $1.8$ \\
\hline
\end{tabular}

Values without an explicit error indicate that the parameter is
unconstrained. Fluxes and luminosities are in the 0.5--10 keV energy
band. Fluxes are absorbed and luminosities are unabsorbed.

$^+$ Even fixing the column density to $1.3\times 10^{21}\cmdue$ we
derive a temperature $0.5$ keV and a reduced $\chi^2=3.2$.

$^*$ Model for a hydrogen atmosphere by Gaensicke, Braje \& Romani
(2002). The temperature collapses to the maximum allowed value of 0.275 keV.

\end{table*}

\newpage

\

\begin{figure*}[!b]
\psfig{figure=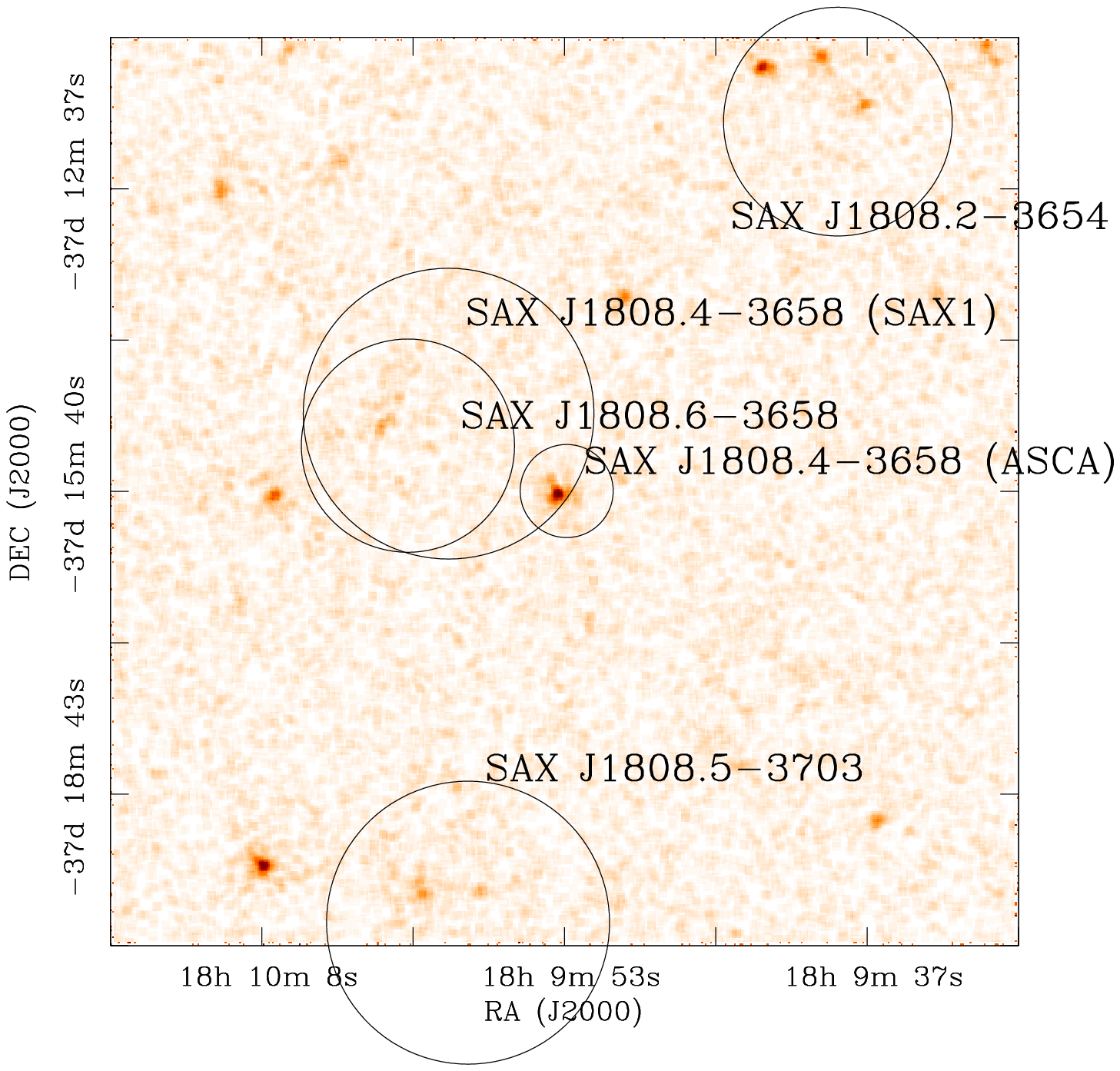,width=10cm}
\psfig{figure=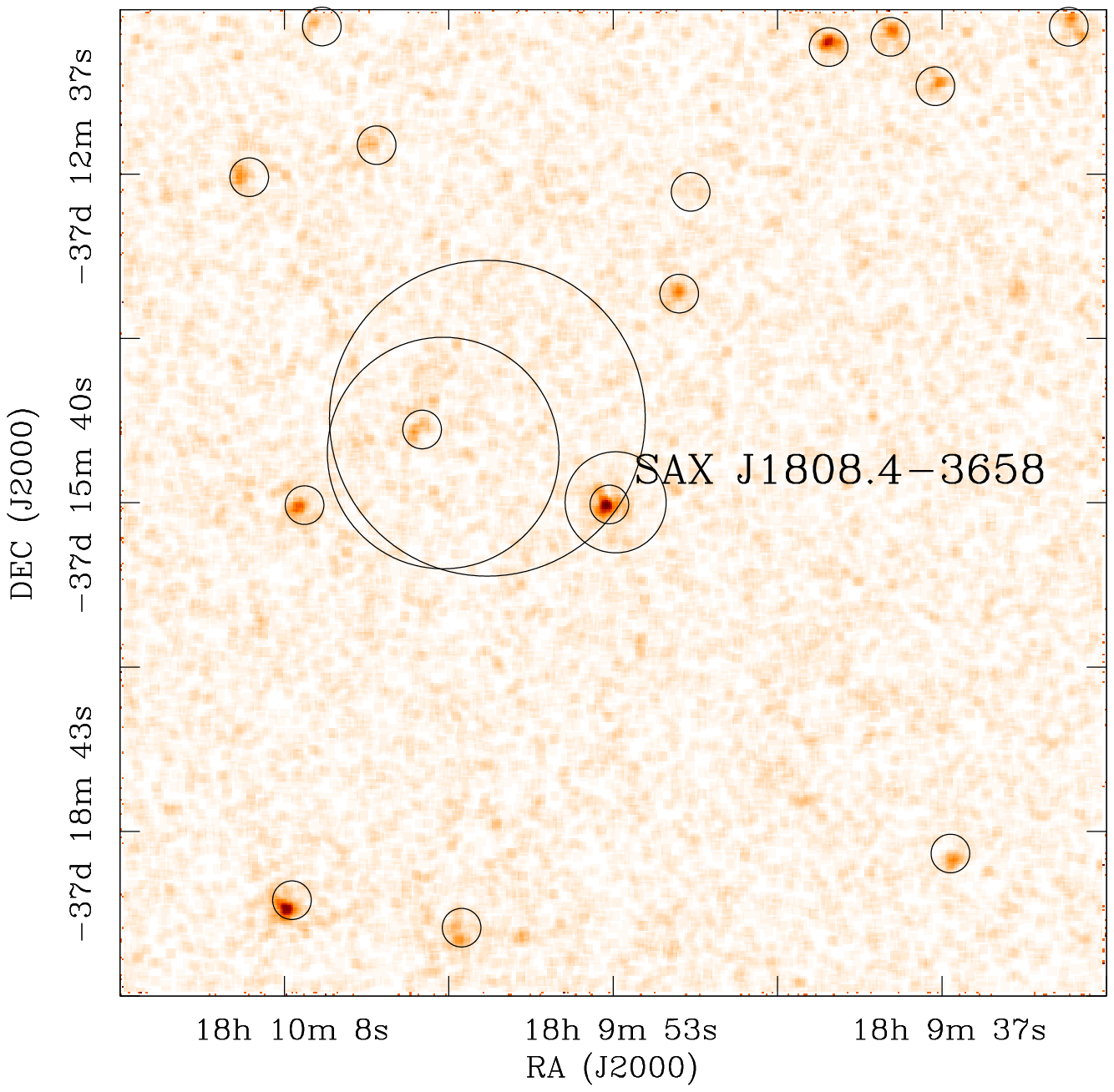,width=10cm}
\caption[h]{Upper panel: XMM-Newton observation of the SAX J1808.4--3658
field with overlaid sources detected during previous observations with
BeppoSAX and ASCA (see text for more details and Table 1).
Lower panel: Same field as above with XMM-Newton detections and 
SAX J1808.4--3658 previous detections. The error circles for the XMM-Newton
sources have been enlarged to $15''$ for display purposes.}
\end{figure*}

\newpage

\begin{figure*}[!t]
\psfig{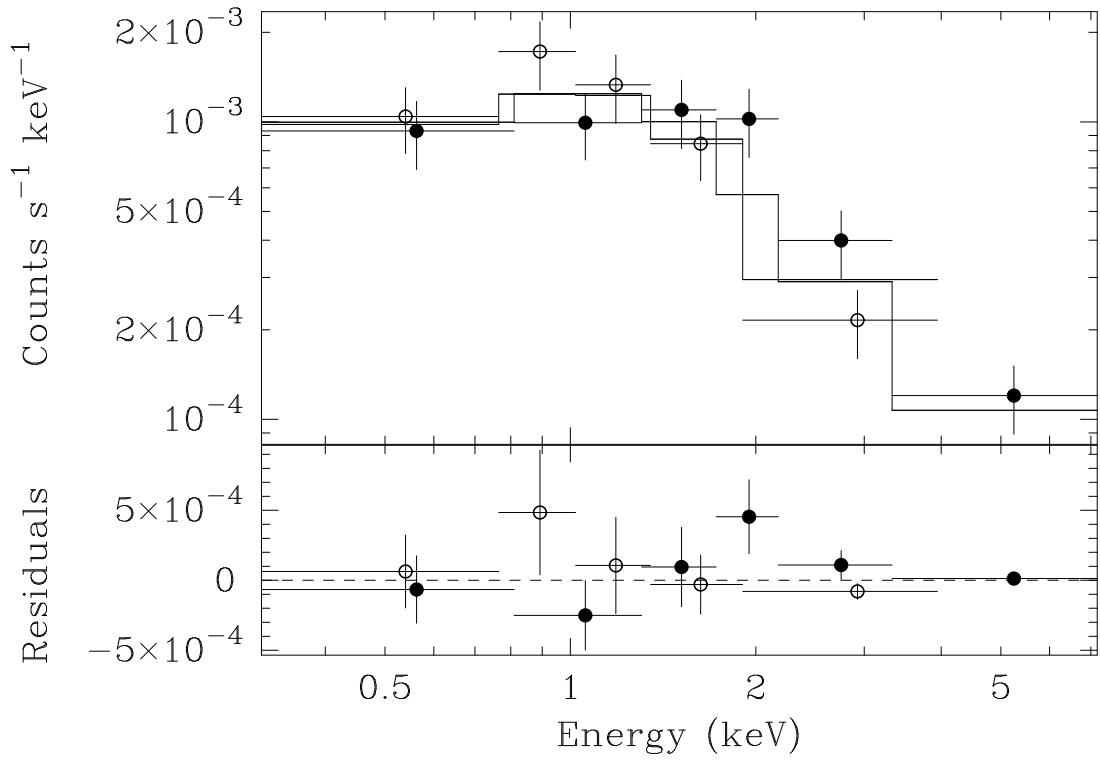}
\caption[h]{MOS1 (open dots) and MOS2 (filled dots) spectrum of SAX J1808.4--3658.
Overlaid is the fit with an absorbed power law model described
in the text. In the lower panel are reported the residuals of the fit.}
\end{figure*}

\newpage
\

\vskip -1truecm
\begin{figure}[!t]
\psfig{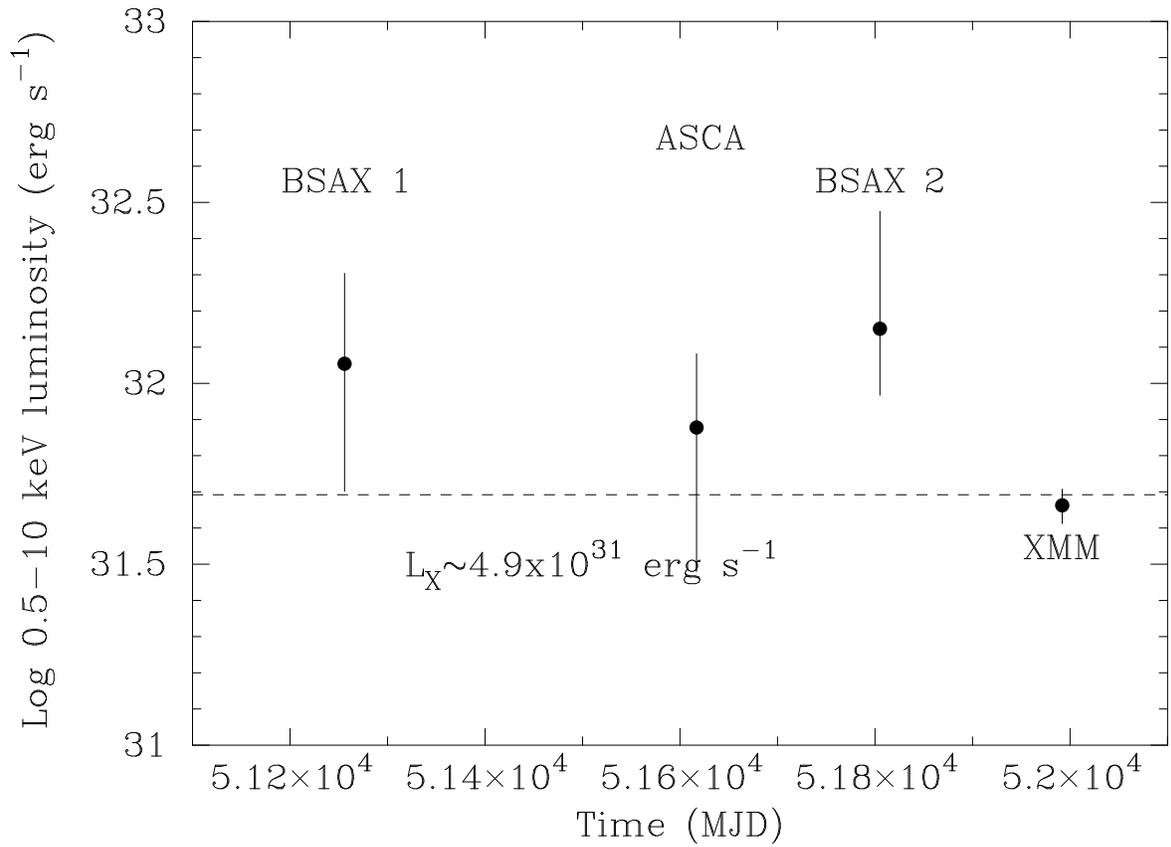}
\vskip -3truecm
\caption[h]{Quiescent 0.5--10 keV unabsorbed luminosities of SAX J1808.4--3658
observed with BeppoSAX, ASCA and XMM-Newton with relative errors
($1\,\sigma$). The dashed line represents the fit with a constant value. In
the case of the BSAX2 observation a smaller uncertainty was reported by
Wijnands et al. (2002b). The small number of counts, the high BeppoSAX
background within the extraction radius (comparable to the source flux), the a 
posteriori discovery of a weak source within this radius make the flux
uncertainty quoted by Wijnands et al. (2002b) too small.}
\end{figure}


\begin{references}

\reference {}
Asai, K., et al. 1996, PASJ 48 257

\reference {}
Asai, K., et al. 1998, PASJ 50 611

\reference {}
Baldi, A. et al. 2002, ApJ 564 190

\reference {}
Bildsten, L., Chakrabarty, D. 2001, ApJ 557 292

\reference {}
Brown, E.F., Bildsten, L., Rutledge, R.E. 1998, ApJ 504 L95

\reference {}
Burderi, L., D'Antona, F., Burgay, M., 2002, ApJ in press (astro-ph/0203387)

\reference {}
Campana, S., Stella, L. 2000, ApJ 541 849

\reference {}
Campana, S., et al. 1997, A\&A 324 941

\reference {}
Campana, S., et al. 1998a, A\&A Rev. 8 279

\reference {}
Campana, S., et al. 1998b, ApJ 499 L65

\reference {}
Campana, S., et al. 2000, A\&A 358 583

\reference {}
Campana, S., et al. 2002, Proc. Symposium ``New Vision of the
X--ray Universe in the XMM-Newton and Chandra Era", ed. F. Jansen (in press)
(2002) (European Space Agency ESA SP-488) 

\reference {}
Chakrabarty, D., Morgan, E.H. 1998, Nat 394 347

\reference {} 
Colpi, M., Geppert, U., Page, D., Possenti, A. 2001, ApJ, 548, L175 

\reference {}
D'Amico, N., et al. 2001, ApJ, 561, L89

\reference {}
Dotani, T., Asai, K., Wijnands, R. 2000, ApJ 543 L145

\reference {}
Ferraro, F., Possenti, A., D'Amico, N., Sabbi, E. 2001, ApJ, 561, L93

\reference {}
Garcia, M.R., et al. 2001, ApJ 553 L47

\reference {}
Gilfanov, M., Revnivtsev, M., Sunyaev, R., Churazov, E. 1998, A\&A, 338, L83

\reference {}
Grindlay, J., Heinke, C.O., Edmonds, P.D., Murray, S.S., Cool, A.M. 2001, ApJ, 
563, L53

\reference {}
Heise, J., et al. 1998, Astrophys. Lett. Comm. 38 301

\reference {}
in't Zand, J.J.M. 2001a, 4th INTEGRAL workshop ``Exploring the gamma-ray
universe", Alicante (Sep. 2000), ESA-SP series in press (astro-ph/0104299)

\reference {}
in't Zand, J.J.M., et al. 2001b, A\&A 372 916

\reference {}
King, A.R. 2000, MNRAS 315 L33

\reference {}
Kirsch, M., 2002, ``EPIC status of calibration and data analysis'',
XMM-SOC-CAL-TN-0018, in preparation 

\reference {}
Perri, M., Capalbi, M. 2002, ``An improvement of
BeppoSAX LECS and MECS source positioning accuracy'', ASDC Tech. Report 
(http://www.asdc.asi.it/bepposax/report/report.html)

\reference {}
Psaltis, D., Chakrabarty, D. 1999, ApJ 521 332

\reference {}
Rutledge, R.E., et al. 2000, ApJ 529 985

\reference {}
Rutledge, R.E., et al. 2002, ApJ in press (astro-ph/0204196)

\reference {}
Stella, L., et al., 1994, ApJ 423 L47

\reference {}
Stella, L., et al., 2000, ApJ 537 L115

\reference {}
Str\"uder, L., et al. 2001, A\&A 365 L18

\reference {}
van der Klis, M. 2000, ARA\&A 38 717

\reference {}
Watson, M.G., et al. 2001, A\&A 365 L51

\reference {}
Wijnands R., van der Klis M., 1998, Nat 394 344

\reference {}
Wijnands R., Guainazzi M., van der Klis M., Mendez M., 2002a, ApJ in press
(astro-ph/0202398) 

\reference {}
Wijnands R., et al. 2002b, ApJ 571 429


\end{references}
\end{document}